\RequirePackage{fix-cm}
\documentclass[twocolumn,epjc3]{svjour3}
\smartqed  % flush right qed marks, e.g. at end of proof
\RequirePackage{graphicx}
\journalname{Eur. Phys. J. C}
\begin{document}

\title{Tetraquark states in the bottom sector and the status of the $Y_b$(10890) state%\thanksref{t1}
}


%\author{Smruti Patel$^1$\\thanksref{{e-mail:} fizixsmriti31@gmail.com}, P C Vinodkumar$^1$\thanks{{e-mail:} p.c.vinodkumar@gmail.com}, and Shashank Bhatnagar$^2$\thanks{{e-mail:} shashank$\_$bhatnagar@yahoo.com}}
\author{Smruti Patel\thanksref{e1} and P C Vinodkumar\thanksref{e2}
       %\and
       %\and
       %\and
       %etc.
%Shashank bhatnagar\thanksref{e2,addr2}
}

\thankstext{e1}{e-mail: fizix.smriti@gmail.com}
\thankstext{e2}{e-mail: p.c.vinodkumar@gmail.com}


\institute{Department of Physics, Sardar Patel University,Vallabh Vidyanagar-388120, Gujarat, India.}

\date{Received: date / Accepted: date}

\maketitle

\begin{abstract}
We have done the exploratory study of bottom tetraquarks ($[bq\bar b \bar q];{q\in u,d}$) in the diquark-\\antidiquark framework with the inclusion of spin hyperfine, spin-orbit and tensor components of the one gluon exchange interaction. Our focus here is on the $Y_b$(10890) and other exotic states in the bottom sector. We have predicted some of the bottom counterparts to the charm tetraquark candidates. Our present study shows that if $Z_b(10610)$ and $Z_b(10650)$ are diquark-diantiquark states then they have to be first radial excitations only and we have predicted $Z_b(10650)$ state as first radial excitation of tetraquark state $X_b$ (10.143-10.230). We have identified  $X_b$ state with $J^{PC}= 1^{+-}/0^{++}$ as being the analogue of $Z_c(3900)$. An observation of the $X_b$ will provide a deeper insight into the exotic hadron spectroscopy and is helpful to unravel the nature of the states connected by the heavy quark symmetry.
We particularly focus on the lowest P wave $[bq][\bar b\bar q]$ states with $J^{PC}=1^{--}$ by computing their leptonic, hadronic and radiative decay widths to predict the status of still controversial $Y_b$(10890) state. Apart from this, we have also shown here the possibility of mixing of P wave states. In the case of mixing of $1^{--}$ state with different spin multiplicities, we found that predicted masses of the  mixed P states differ from $Y_b$(10890) state only by $\pm20$ MeV energy difference which can be helpful to resolve further the structure of $Y_b$(10890).
\end{abstract}
\section{Introduction}
\label{intro}
A plethora of new kind of states which have been observed recently has inspired extensive interest in revealing the underlying structure of these newly observed states. Exploration of these states will improve our understanding of non-perturbative
QCD. In recent years a significant experimental progress has been achieved regarding  discoveries of bottomonium-like and charmonium-like charged manifestly exotic resonances $Z_b$(10610), $Z_b$(10650) \cite{KF,Karliner,bonder,I,PK,babar,Garmash}, $Z_c$(3900) \cite{MA1,liu,TX,MA2,MA3} and $Z_c$(4020/4025) \cite{MA4,MA5,Chilikin,MA6}.Their production mechanism and decay rates are not compatible with a standard quarkonium interpretation. A huge effort in understanding the nature of these new states and in building a new spectroscopy is forthcoming.

In the recent years strong experimental evidence from B and charm factories has been accumulating for the existence of exotic new quarkonia states, narrow resonances called X, Y, Z particles which do not seem to have a simple $q\bar q$ structure. Their masses and decay modes show that they contain a heavy quark-antiquark pair, but their quantum numbers are such that
they must also contain a light quark-antiquark pair\cite{marek}. The theoretical challenge has been to determine the nature of these resonances. Their production mechanism, masses, decay widths, spin-parity assignments and decay modes have been revisited recently \cite{Dias,VR,Hong}. The term exotica labels states which have an identical number of quarks and antiquarks but defy an ordinary meson classification. Many exotic states in the charm sector with $c\bar c$ content have been discovered by Belle and others \cite{MA1,zupanc}. While there are most likely many more which are yet unknown and many of them should also reflect in the $b\bar b$ sector according to heavy quark symmetry. The non-discovery of the respective $b\bar b$ partners of the charmonium-like exotica would be even more enigmatic. Belle collaboration has extended the study of the XYZ exotic state family to the bottomonium sector by claiming the observations of two exotica states in $\Upsilon(5S)$ decays \cite{bonder}. The CMS experiment also searched for the bottomonium partner of $X(3872)$ at hadron colliders \cite{CMS} in the $\Upsilon(1S)\pi\pi$ decay mode and found no evidence for the $X_b$ state while the ratio of the cross section $X_b$ to $\Upsilon(2S)$ shows upper limit in the range of (0.9-5.4)$\%$ at 95$\%$ confidence level for $X_b$ masses between 10-11 GeV. Those are the first upper limits on the production of a possible $X_b$ state at a hadron collider.
 Currently there are pending, unanswered questions concerning the exotic spectroscopy in the heavy quark sectors especially in the bottom sector. To promote the endeavor of understanding the heavy exotic states, the exploration of the bottom sector is important. Motivated by the BaBar's discovery of large $Y(4260) \rightarrow\pi+\pi+J/\psi$ signal discovered in the charmonium mass region, Belle experiment have searched for similar state in the bottomonium sector\cite{Hou}.
They observed partial decay widths  $\Upsilon(5S) \rightarrow \pi+\pi+\Upsilon(nS)$ (n = 1, 2, 3)  associated with the peak in the $\pi+\pi+\Upsilon(nS)$ cross section hundreds of times larger than the theoretical predictions \cite{KF} and the corresponding measured rates for the $\Upsilon(4S)$\cite{PDG12}. This observation suggests the presence of a new, non-conventional hadronic state in the bottom sector equivalent of the $Y(4260)$ of the charm sector with mass around 10.890 GeV \cite{ali} which is referred  as $Y_b(10890)$ state. Indeed, there exist three candidates up to date, namely the states labeled $Y_b$(10890), $Z_b$(10610) and $Z_b$(10650), observed by Belle [3]. Not only new states are waiting to be discovered but also the existence of $Y_b$(10890) needs to be established or refuted. The  $Y_b$(10890) is a potential exotic state still remains to be confirmed since its observation first reported by the Belle collaboration \cite{KF,chen}. Looking into the interest in this case, present study is particularly focus on the negative parity $1^{--}$ exotic states. Apart from its spin parity, study of its di-leptonic, hadronic and radiative decay widths also help us to solve the puzzling features of this state. The interpretation of the hidden  bottom four quark state as a tetraquark exotic states has been advanced and has been studied in considerable detail \cite{EB1,ali,AA,A1,A2,A3,brink,semay,lattice,wang,PB}. The experimental search for tetraquark states is a very difficult problem, since exotic candidates are nothing but the resonances immersed in the excited
hadron spectra and moreover they usually decay to several hadrons.  Their mass and decay products put them in the category of  quarkonia-like resonances but their masses do not fit into the conventional quark model spectrum of quark-antiquark mesons \cite{sg,PB1}. However, to confirm a new resonance it is necessary to study all its properties with high level of accuracy including its mass and width. In this work, we develop phenomenology to study some of the theoretical problems of multiquarks and predict multiquark bound states and resonances. In particular, as a benchmark, we study in detail the heavy-light antilight-antiheavy systems who are expected to produce tetraquarks. Despite the intense experimental attempts, these resonances are still mysterious and complicated and we still lack of a comprehensive theoretical framework. In particular, the most popular phenomenological models proposed to explain the internal structure of these particles are the compact tetraquark in the constituent diquark-antidiquark picture and the loosely bound di-meson molecular picture. Following Gell-Mann$'$s suggestion of the possibility of diquark stucture \cite{gellmann}, various authors have introduced effective degrees of freedom of diquarks in order to describe tetraquarks as composed of a constituent diquark and diantiquark  using QCD sum rules \cite{1,2}. This concept of diquark was even used to account for some experimental phenomena \cite{RLJ}. The authors of refs \cite{Drenska,Drenska1} studied the tetraquark systems in the diquark-antidiquark picture using the chromo-magnetic interactions. In the same way Maiani et al. \cite{LM,LFAD,LM1} also studied tetraquarks and pentaquarks systems by considering this concept of diquark. In their study they have included the  spin– spin interactions. On the other hand Ebert et. al. \cite{EB,EB1,EB2} employed the relativistic quark model based on the quasi-potential approach in order to find the mass spectra of hidden heavy tetraquark systems. Unlike Maiani et al., they ignored the spin$-$spin interactions inside the diquark and anti-diquark. The presence of a coherent diquark structure within tetraquarks helps us to treat the problem of four-body to that of two two-body interactions. In the present case, we employ the diquark and anti-diquark picture in the beauty sector and compute the mass spectra of the diquark-antidiquarks $[bq\bar b \bar q];{q\in u,d}$ in the ground and orbitally excited states with the inclusion of both $S=0$ and $S=1$ diquarks. We present the formalism of the study of hidden bottom tetraquark states in section 2. In section 3, we discuss the  $Y_b$ states and their decay properties. We conclude and discuss our findings in section 4.
\section{Theoretical framework}
\label{sec:1}
In this paper we shall take a different path and investigate different ways in which the experimental data can
be reproduced. We have treated the four particle system as two-two body systems interacting through effective potential of the same form of the two body interaction potential of Eq.(1). The existence of exotic hadrons of the diquarks-diantiquarks pair called tetraquarks or diquakonia is a problem which was foremost raised about 20 years ago and was used to describe scalar mesons below 1GeV in 1977 by R. Jaffe\cite{jaffe,j1}. He suggested the idea of strongly correlated two-quarks-two-antiquarks states to baryon-antibaryon channels where the MIT bag model used to predict the quantum numbers and the masses of prominent states. There are two types of diquarks one is $S=0 $ good (scalar) diquarks and another one is $S=1$ bad (vector) diquarks. We have available lattice results which favors the evidence of an attractive diquark(antidiquark) channel for the good diquarks (color antitriplet, flavour antisymmetric) with spin $S=0$ in accordance with Jaffe's proposal. On the other hand there is no lattice results available for an attractive channel for the bad diquarks i.e. with spin $S=1$. Here, we use the fact the effective QCD-lagrangian is independent of spin in the heavy quark limit and we incorporate the diquark with $S=1$ also in computing the mass spectra. There are many methods to estimate the mass of a hadron, among which phenomenological potential model is a fairly reliable one especially for heavy hadrons \cite{PRC,BP,sam,DP}.\\
In the present study, the non-relativistic interaction potential we have used is the Cornell potential which consists of a central term V (r) which is just the sum of the Coulomb (vector) and linear
confining (scalar) parts given by

\begin{equation}\label{eq:1}
V(r)=V_V+V_S=k_s\frac{\alpha_s}{r}+Ar+B
\end{equation}

\begin{eqnarray}
k_s&=&-4/3 \ \ for \ \ q\bar q\ \ \ \nonumber \\
   &=&-2/3 \ \ for \ \ qq \ \ or \ \ \bar q\bar q
\end{eqnarray}

The value of the $\alpha_s$, the running coupling constant is determined by \cite{ebert}
\begin{equation}\label{eq:2}
\alpha_s(\mu^2)=\frac{4\pi}{(11-\frac{2}{3}n_f)(\ln\frac{\mu^2+{M^2_B}}{\Lambda^2})}
\end{equation}

where $\mu$ = $2m_am_b/(m_a + m_b)$, $\Lambda =0.413$ GeV, $M_B$ is the background mass
and $n_f$ is number of flavours \cite{ebert}. The model parameters we have used in the present study are same as in refs\cite{ebert,ebert1}. The constituent quark masses employed here are: $m_u=m_d=0.33 GeV$ and $m_b=4.88 GeV$.
The degeneracy of these exotic states are removed by including the spin-dependent part of the usual one gluon exchange potential \cite{Barnes,Olga,Voloshin,Eichten}. The potential description extended to spin-dependent interactions results in three types of interaction terms such as the spin-spin, the spin-orbit and the tensor part. Accordingly, the spin dependent part $V_{SD}$ is given by
\begin{eqnarray}\label{eq:3}
V_{SD}&=&V_{SS}\left[\frac{1}{2}(S(S+1)-\frac{3}{2}))\right] \nonumber \\ && +V_{LS}\left[\frac{1}{2}(J(J+1)-S(S+1)-L(L+1))\right] \nonumber \\ && +V_{T}\left[12\left(\frac{(S_1.r)(S_2.r)}{r^2}-\frac{1}{3}(S_1.S_2)\right)\right]
\end{eqnarray}

The coefficient of these spin dependent terms of Eq.(\ref{eq:3}) can be written in terms of the vector ($V_V$ ) and scalar ($V_S$ ) parts of the static potential described in Eq.(1) as

\begin{equation}\label{eq:4}
V^{ij}_{LS}(r)=\frac{1}{2M_iM_jr}\left[ 3\frac{dV_V}{dr}-\frac{dV_S}{dr}\right]
\end{equation}

\begin{equation}\label{eq:5}
V^{ij}_T(r)=\frac{1}{6M_iM_j}\left[ 3\frac{d^2V_V}{dr^2}-\frac{1}{r}\frac{dV_S}{dr}\right]
\end{equation}

\begin{equation}\label{eq:6}
V^{ij}_{SS}(r)=\frac{1}{3M_iM_j}\nabla^2 V_V=\frac{16\pi\alpha_s}{9M_iM_j}\delta^3(r)
\end{equation}

Where $M_i$, $M_j$ correspond to the masses of the respective constituting two-body systems. The Schr\"{o}dinger equation with the potential given by
Eq. (1) is numerically solved using the Mathematica notebook of the Runge-Kutta method \cite{Lucha} to
obtain the energy eigen values and the corresponding wave functions.
\begin{table*}
\begin{center}
\caption{Mass spectra of four quark states in the diquark-antidiquark picture (For $L_1=0$, $L_2=0$)(in GeV))}. \label{tab1}
\begin{tabular}{cccccccccccccc}
\hline\\
$S_d$	&	$L_d$	&	$S_{\bar d}$	&	$L_{\bar d}$ &	$J_d$	&	$J_{\bar d}$	&	$J$	&	$J^{PC}$  	&	$^{2s+1}X_J$	&	$M_{cw}$	&	$V_{SS}$	&	$V_{LS}$	&	$V_{T}$	&	$M_{J}$	\\
\hline
	&		&		&		&		&		&		&		&		&		&		&		&		&		\\
0	&	0	&	0	&	0	&	0	&	0	&	0	&	$0^{++}$	&	$^1S_0$	&	10.309	&	0.0	&	0.0	&	0.0	&	10.309	\\
	&		&		&		&		&		&		&		&		&		&		&		&		&		\\
1	&	0	&	0	&	0	&	1	&	0	&	1	&	$1^{+-}$	&	$^3S_1$	&	10.316	&	0.0	&	0.0	&	0.0	&	10.316	\\
	&		&		&		&		&		&		&		&		&		&		&		&		&		\\
1	&	0	&	1	&	0	&	1	&	1	&	0	&	$0^{++}$	&	$^1S_0$	&	10.323	&	-0.179	&	0.0	&	0.0	&	10.143	\\
	&		&		&		&		&		&	1	&	$1^{+-}$	&	$^3S_1$	&	10.323	&	-0.089	&		&		&	10.233	\\
	&		&		&		&		&		&	2	&	$2^{++}$	&	$^5S_1$	&	10.323	&	0.089	&		&		&	10.413	\\

\hline

\end{tabular}

\end{center}

\end{table*}

\subsection{The four-quark state in diquark-antidiquark picture}
\label{sec:2.1}
In this section, we calculate the mass spectra of tetra-\\quarks with hidden bottom as the bound states of two clusters ($Qq$ and $\bar Q\bar q$), (Q = b; q = u, d). We think of the diquarks as two correlated quarks with no internal spatial excitation. Because a pair of quarks cannot be a color singlet, the diquark can only be found confined into hadrons and used as effective degree of freedom. Heavy light diquarks can be the building blocks of a rich spectrum of exotic states which can not be fitted in the conventional quarkonium assignment. Maiani et al \cite{LM} in the framework of the phenomenological constituent quark model considered the masses of hidden/open charm diquark-antidiquark states in terms of the constituent diquark masses with their spin-spin interactions included. We discuss the spectra in the framework of a non-relativistic hamiltonian including chromo-magnetic spin-spin interactions between the \\quarks (antiquarks) within a diquark(antidiquark. Masses of diquark (antidiquark) states are obtained by numerically solving the  Schr\"{o}dinger equation with the respective two body potential given by Eq.(1) and incorporating the respective spin interactions described by Eq.(\ref{eq:3}) perturbatively.\\
 In the diquark-antidiquark structure, the masses of the  diquark/diantiquark system are given by:

 \begin{equation}\label{eq:7}
m_d=m_{Q}+m_{q}+E_{d}+{\langle V_{SD}\rangle_{Qq}}
\end{equation}

\begin{equation}\label{eq:8}
m_{\bar d}=m_{\bar Q}+m_{\bar q}+E_{\bar d}+{\langle V_{SD}\rangle_{\bar Q\bar q}}
\end{equation}
Further, the same procedure is adopted to compute the binding energy of the diquark-antidiquark  bound system as
\begin{equation}\label{eq:9}
M_{d-\bar d}=m_d+m_{\bar d}+E_{d\bar d}+{\langle V_{SD}\rangle}_{d\bar d}
\end{equation}

Where $Q$ and $q$ represents the heavy quark and light quark respectively. In the present paper, $d$ and $\bar d$ represents diquark and antidiquark respectively. While $E_d$, $E_{\bar d}$, $E_{d \bar d}$ are the energy eigen values of the diquark, antidiquark and diquark-antidiquark system respectively. The spin-dependent potential ($V_{SD}$) part of the hamiltonian described by Eq.(\ref{eq:3}) has been treated perturbatively. Details of the computed results are listed in Table 1, 2, 3, 4 and 5 for the low lying positive parity and negative parity states respectively.\\
%%%%%%%%%%%%%%%%%%%%%%%%%%%%%%%%%%%%%%%%%%%%%%%%%%%%%%%%%%%%%%%%%%%%%%%%%%%%%%%%%%%%%%%%%%%%%%%%%%%%%%%%%%%%%%%%%%%%%%%%%%%%%

%%%%%%%%%%%%%%%%%%%%%%%%%%%%%%%%%%%%%%%%%%%%%%%%%%%%%%%%%%%%%%%%%%%%%%%%%%%%%%%%%%%%%%%%%%%%%%%%%%%%%%%%%%%%%%%%%%%%%%%%%%%%%%%%%
\begin{table*}
\begin{center}
\caption{Mass spectra of four quark states in the diquark-antidiquark picture($L_1=1$, $L_2=0$)(in GeV))}. \label{tab1}
\begin{tabular}{cccccccccccccc}
\hline\\
$S_d$	&	$L_d$	&	$S_{\bar d}$	&	$L_{\bar d}$ &	$J_d$	&	$J_{\bar d}$	&	$J$	&	$J^{PC}$  	&	$^{2s+1}X_J$	&	$M_{cw}$	&	$V_{SS}$	&	$V_{LS}$	&	$V_{T}$	&	$M_{J}$	\\
\hline
	&		&		&		&		&		&		&		&		&		&		&		&		&		\\
0	&	1	&	0	&	0	&	1	&	0	&	1	&	$1^{- -}$	&	$^1P_1$	&	10.917	&	0.0	&	0.0	&	0.014	&	10.931	\\
	&		&		&		&		&		&		&		&		&		&		&		&		&		\\
1	&	1	&	0	&	0	&	0	&	0	&	0	&	$0^{- +}$	&	$^3P_0$	&	10.917	&	0.000	&	-0.0059	&	-0.0286	&	10.883	\\
	&		&		&		&	1	&		&	1	&	$1^{- +}$	&	$^3P_1$	&		&	0.000	&	-0.0029	&	-0.011	&	10.921	\\
	&		&		&		&	2	&		&	2	&	$2^{- +}$	&	$^3P_2$	&		&	0.000	&	0.0029	&	-0.0256	&	10.913	\\
	&		&		&		&		&		&		&		&		&		&		&		&		&		\\
1	&	1	&	1	&	0	&	0	&	1	&	1	&	$1^{- -}$	&	$^1P_1$	&	10.925	&	-0.019	&	0.0	&	-0.0233	&	10.882	\\
	&		&		&		&		&		&		&		&		&		&		&		&		&		\\
	&		&		&		&	1	&	1	&	0	&	$0^{- +}$	&	$^3P_0$	&		&	-0.0095	&	-0.0059	&	-0.0467	&	10.862	\\
	&		&		&		&		&		&	1	&	$1^{- +}$	&	$^3P_1$	&		&		&	-0.0029	&	-0.011	&	10.900	\\
	&		&		&		&		&		&	2	&	$2^{- +}$	&	$^3P_2$	&		&		&	0.0029	&	-0.026	&	10.892	\\
	&		&		&		&		&		&		&		&		&		&		&		&		&		\\
	&		&		&		&	2	&	1	&	1	&	$1^{- -}$	&	$^5P_1$	&		&	0.0095	&	-0.0088	&	-0.072	&	10.853	\\
	&		&		&		&		&		&	2	&	$2^{- -}$	&	$^5P_2$	&		&		&	-0.0029	&	0.0256	&	10.957	\\
	&		&		&		&		&		&	3	&	$3^{- -}$	&	$^5P_3$	&		&		&	0.006	&	-0.037	&	10.903	\\

\hline

\end{tabular}

\end{center}

\end{table*}
%%%%%%%%%%%%%%%%%%%%%%%%%%%%%%%%%%%%%%%%%%%%%%%%%%%%%%%%%%%%%%%%%%%%%%%%%%%%%%%%%%%%%%%%%%%%%%%%%%%%%%%%%%%%%%%%%%%%%%%%%%%%%%%
\begin{figure}
%\begin{centering}

\resizebox{0.55\textwidth}{!}{%
  \includegraphics{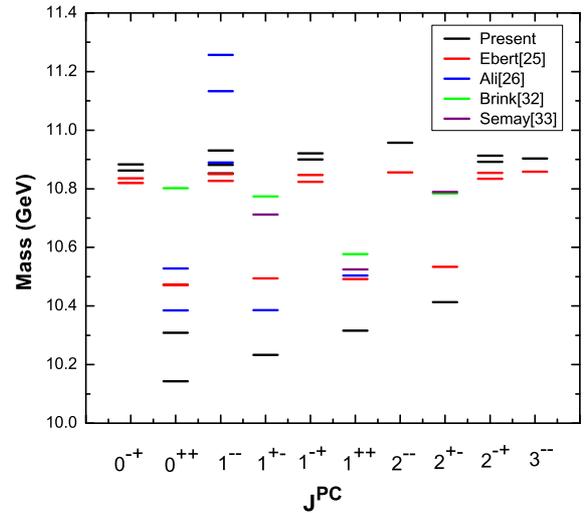}
}
%\vspace{5cm}       % Give the correct figure height in cm
\caption{ Mass spectra of Bottom tetraqurk states(in GeV).}
\label{fig:1}       % Give a unique label
%\end{centering}
\end{figure}
%%%%%%%%%%%%%%%%%%%%%%%%%%%%%%%%%%%%%%%%%%%%%%%%%%%%%%%%%%%%%%%%%%%%%%%%%%%%%%%%%%%%
\begin{table*}
\begin{center}
\caption{Mass spectra of four quark states in the diquark-antidiquark picture($L_1=1$, $L_2=1$)(in GeV))}. \label{tab1}
\begin{tabular}{cccccccccccccc}
\hline\\
$S_{d}$	&	$L_{d}$	&	$S_{\bar d}$	&	$L_{\bar d}$ &	$J_{d}$	&	$J_{\bar d}$	&	$J$	&	$J^{PC}$  	&	$^{2s+1}X_J$	&	$M_{cw}$	&	$V_{SS}$	&	$V_{LS}$	&	$V_{T}$	&	$M_{J}$	\\
\hline

0	&	1	&	0	&	1	&	1	&	1	&	0	&	$0^{++}$	&	$^1S_0$	&	10.843	&	0.0	&	0.0	&	0.0	&	10.843	\\
	&		&		&		&		&		&	1	&	$1^{+-}$	&	$^1P_1$	&	11.187	&	0.0	&	0.0	&	0.0139	&	11.201	\\
	&		&		&		&		&		&	2	&	$2^{++}$    &   $^1D_2$	&	11.348	&	0.0	&	0.0	&	0.002	&	11.350	\\
1	&	1	&	0	&	1	&	0	&	1	&	1	&	$1^{+-}$	&	$^3S_1$	&	10.843	&	0.0	&	0.0	&	0.0	&	10.843	\\
	&		&		&		&		&		&		&		&		&		&		&		&		&		\\
	&		&		&		&		&		&	0	&	$0^{++}$	&	$^3P_0$	&	11.188	&	0.0	&	-0.0059	&	-0.0278	&	11.154	\\
	&		&		&		&	1	&	1	&	1	&	$1^{+-}$	&	$^3P_1$	&		&	0.0	&	-0.0029	&	0.0069	&	11.192	\\
	&		&		&		&		&		&	2	&	$2^{++}$	&	$^3P_2$	&		&	0.0	&	0.0029	&	-0.0069	&	11.184	\\
	&		&		&		&		&		&		&		&		&		&		&		&		&		\\
	&		&		&		&		&		&	1	&	$1^{+-}$	&	$^3D_1$	&	11.348	&	0.0	&	-0.0007	&	-0.003	&	11.344	\\
	&		&		&		&	2	&	1	&	2	&	$2^{++}$	&	$^3D_2$	&		&	0.0	&	-0.00024	&	0.0015	&	11.350	\\
	&		&		&		&		&		&	3	&	$3^{+-}$	&	$^3D_3$	&		&	0.0	&	0.00049	&	-0.00135	&	11.347	\\
	&		&		&		&		&		&		&		&		&		&		&		&		&		\\
1	&	1	&	1	&	1	&	0	&	0	&	0	&	$0^{++}$	&	$^1S_0$	&	10.843	&	-0.174	&	0.0	&	0.0	&	10.640	\\
	&		&		&		&	1	&		&	1	&	$1^{+-}$	&	$^3S_1$	&		&	-0.092	&	0.0	&	0.0	&	10.751	\\
	&		&		&		&	2	&		&	2	&	$2^{++}$	&	$^5S_2$	&		&	0.092	&	0.0	&	0.0	&	10.936	\\
	&		&		&		&		&		&		&		&		&		&		&		&		&		\\
	&		&		&		&	0	&	1	&	1	&	$1^{+-}$	&	$^1P_1$	&	11.188	&	-0.019	&	0.0	&	-0.023	&	11.145	\\
	&		&		&		&	1	&	1	&	0	&	$0^{++}$	&	$^3P_0$	&		&	-0.0098	&	-0.0059	&	-0.046	&	11.126	\\
	&		&		&		&		&		&	1	&	$1^{+-}$	&	$^3P_1$	&		&	-0.0098	&	-0.0029	&	-0.011	&	11.163	\\
	&		&		&		&		&		&	2	&	$2^{++}$	&	$^3P_2$	&		&	-0.0098	&	0.0029	&	-0.025	&	11.155	\\
	&		&		&		&	2	&	1	&	1	&	$1^{+-}$	&	$^5P_1$	&		&	0.0098	&	-0.0088	&	-0.071	&	11.117	\\
	&		&		&		&		&		&	2	&	$2^{++}$	&	$^5P_2$	&		&	0.0098	&	-0.0029	&	0.0255	&	11.220	\\
	&		&		&		&		&		&	3	&	$3^{+-}$	&	$^5P_3$	&		&	0.0098	&	0.0059	&	-0.0371	&	11.167	\\
	&		&		&		&		&		&		&		&		&		&		&		&		&		\\
	&		&		&		&	0	&	2	&	2	&	$2^{++}$	&	$^1D_2$	&	11.348	&	-0.0072	&	0.0	&	-0.0033	&	11.338	\\
	&		&		&		&	1	&	2	&	1	&	$1^{+-}$	&	$^3D_1$	&		&	-0.0036	&	0.007	&	-0.0057	&	11.339	\\
	&		&		&		&		&		&	2	&	$2^{++}$	&	$^3D_2$	&		&	-0.0036	&	-0.0024	&	-0.0017	&	11.344	\\
	&		&		&		&		&		&	3	&	$3^{+-}$	&	$^3D_3$	&		&	-0.0036	&	0.00049	&	-0.004	&	11.341	\\
	&		&		&		&		&		&	0	&	$0^{++}$	&	$^5D_0$	&		&	0.0036	&	-0.0014	&	-0.017	&	11.333	\\
	&		&		&		&	2	&	2	&	1	&	$1^{++}$	&	$^5D_1$	&		&	0.0036	&	-0.0012	&	-0.010	&	11.340	\\
	&		&		&		&		&		&	2	&	$2^{++}$	&	$^5D_2$	&		&	0.0036	&	-0.0007	&	-0.0033	&	11.351	\\
	&		&		&		&		&		&	3	&	$3^{+-}$	&	$^5D_3$	&		&	0.0036	&	0	&	0.0047	&	11.357	\\
	&		&		&		&		&		&	4	&	$4^{++}$	&	$^5D_4$	&		&	0.0036	&	0.0009	&	-0.0074	&	11.346	\\

\hline

\end{tabular}

\end{center}

\end{table*}

\begin{table*}
\begin{center}
\caption{$1^{st}$ radially excited mass spectra of four quark states in the diquark-antidiquark picture(For $L_1=0$, $L_2=0$)(in GeV))}. \label{tab8}
\begin{tabular}{cccccccccccccc}
\hline\\
$S_d$	&	$L_d$	&	$S_{\bar d}$	&	$L_{\bar d}$ &	$J_d$	&	$J_{\bar d}$	&	$J$	&	$J^{PC}$  	&	$^{2s+1}X_J$	&	$M_{cw}$	&	$V_{SS}$	&	$V_{LS}$	&	$V_{T}$	&	$M_{J}$	\\
\hline
	&		&		&		&		&		&		&		&		&		&		&		&		&		\\
0	&	0	&	0	&	0	&	0	&	0	&	0	&	$0^{++}$	&	$^1S_0$	&	10.702	&	0.0	&	0.0	&	0.0	&	10.702	\\
	&		&		&		&		&		&		&		&		&		&		&		&		&		\\
1	&	0	&	0	&	0	&	1	&	0	&	1	&	$1^{+-}$	&	$^3S_1$	&	10.709	&	0.0	&	0.0	&	0.0	&	10.709	\\
	&		&		&		&		&		&		&		&		&		&		&		&		&		\\
1	&	0	&	1	&	0	&	1	&	1	&	0	&	$0^{++}$	&	$^1S_0$	&	10.716	&	-0.066	&	0.0	&	0.0	&	10.650	\\
	&		&		&		&		&		&	1	&	$1^{+-}$	&	$^3S_1$	&	10.716	&	-0.033	&		&		&	10.683	\\
	&		&		&		&		&		&	2	&	$2^{++}$	&	$^5S_1$	&	10.716	&	0.033	&		&		&	10.750	\\

\hline

\end{tabular}

\end{center}

\end{table*}

%%%%%%%%%%%%%%%%%%%%%%%%%%%%%%%%%%%%%%%%%%%%%%%%%%%%%%%%%%%%%%%%%%%%%%%%%%%%%%%%%%%%%%%%%%%%%%%%%%%%%%%%%%%%%%%%%%%%%%%%%%%%%%%%%
\begin{table*}
\begin{center}
\caption{$1^{st}$ radially excited mass spectra of four quark states in the diquark-antidiquark picture($L_1=1$, $L_2=0$)(in GeV))}. \label{tab9}
\begin{tabular}{cccccccccccccc}
\hline\\
$S_d$	&	$L_d$	&	$S_{\bar d}$	&	$L_{\bar d}$ &	$J_d$	&	$J_{\bar d}$	&	$J$	&	$J^{PC}$  	&	$^{2s+1}X_J$	&	$M_{cw}$	&	$V_{SS}$	&	$V_{LS}$	&	$V_{T}$	&	$M_{J}$	\\
\hline
	&		&		&		&		&		&		&		&		&		&		&		&		&		\\
0	&	1	&	0	&	0	&	1	&	0	&	1	&	$1^{- -}$	&	$^1P_1$	&	11.140	&	0.0	&	0.0	&0.011&	11.151	\\
	&		&		&		&		&		&		&		&		&		&		&		&		&		\\
1	&	1	&	0	&	0	&	0	&	0	&	0	&	$0^{- +}$	&	$^3P_0$	&	11.140	&0.000&-0.0047&-0.022&	11.114	\\
	&		&		&		&	1	&		&	1	&	$1^{- +}$	&	$^3P_1$	&&	0.000	&	-0.0023	&	-0.0055	&11.144	\\
	&		&		&		&	2	&		&	2	&	$2^{- +}$	&	$^3P_2$	&		&	0.000	&	0.0023&-0.0055&	11.137	\\
	&		&		&		&		&		&		&		&		&		&		&		&		&		\\
1	&	1	&	1	&	0	&	0	&	1	&	1	&	$1^{- -}$	&	$^1P_1$	&	11.148	&	-0.021	&	0.0	&	-0.018	&	11.108	\\
	&		&		&		&		&		&		&		&		&		&		&		&		&		\\
	&		&		&		&	1	&	1	&	0	&	$0^{- +}$	&	$^3P_0$	&		&	-0.0108	&-0.0047&-0.036	&11.095	\\
	&		&		&		&		&		&	1	&	$1^{- +}$	&	$^3P_1$	&		&		&	-0.0023	&	-0.0092	&	11.125	\\
	&		&		&		&		&		&	2	&	$2^{- +}$	&	$^3P_2$	&		&		&	0.0023	&	-0.010	&	11.119	\\
	&		&		&		&		&		&		&		&		&		&		&		&		&		\\
	&		&		&		&	2	&	1	&	1	&	$1^{- -}$	&	$^5P_1$	&		&	0.0108	&	-0.007	&	-0.057	&	11.094	\\
	&		&		&		&		&		&	2	&	$2^{- -}$	&	$^5P_2$	&		&		&	-0.002	&	0.020	&	11.177	\\
	&		&		&		&		&		&	3	&	$3^{- -}$	&	$^5P_3$	&		&		&	0.004	&	-0.029	&	11.134	\\

\hline

\end{tabular}

\end{center}

\end{table*}
%%%%%%%%%%%%%%%%%%%%%%%%%%%%%%%%%%%%%%%%%%%%%%%%%%%%%%%%%%%%%%%%%%%%%%%%%%%%%%%%%%%%%%%%%%%%%%%%%%%%%%%%%%%%55

%%%%%%%%%%%%%%%%%%%%%%%%%%%%%%%%%%%%%%%%%%%%%%%%%%%%%%%%%%%%%%%%%%%%%%%%%%%%%%%%%%%%%%%%%%%%%%%%%%%%%%%%%%%%%%%%%%%%%%%%%%%%%%%5

\subsection{Mixing of P wave states}
\label{sec:2.2}
In the limit of heavy quark, the spin of the light and heavy degrees of freedom are separately conserved by the strong interaction. So hadrons containing a heavy quark can be simultaneously assigned the quantum numbers $S_{Q\bar q}$, $m_{Q\bar q}$, $\bar Qq$, $m_{\bar Qq}$. Since dynamics depends only on spin of the light degrees of freedom, the hadron will appear in degenerate multiplets of total spin S that can be formed from diquark and antidiquark and accordingly we can classify the states in the convenient way. In the present study, we find that the masses of orbitally excited state with relative angular momentum $L=1$ and total spin $S=0, 1, 2$ corresponding to $^1P_1$, $^3P_1$, $^5P_1$ and $^3P_0$ are close to each other in the mass region around 10.850-11.201GeV. The importance of the linear combination of scalar and axial vector states was noted by Rosner \cite{p1} and he emphasized in the context of the constituent quark model the individual conservation of heavy and light degree of freedom in the heavy quark systems. In the mass spectra shown in Table 2, two $^1P_1$ states with masses 10.931 GeV and 10.882 GeV and another $^5P_1$ state with mass 10.853 GeV are there. Similarly, in the mass spectra shown in Table 3, there are two $^1P_1$ states with masses 10.201 GeV and 10.145 GeV respectively, $^3P_1$ state with mass 11.163 GeV and $^5P_1$ state with mass 11.117 GeV. \\
Generally mixing is done through
\[
\left(
\begin{array}{c}
|P_J>\\
|{P_J}'>
\end{array}
\right)
=U^{-1}\left(
\begin{array}{c}
|\alpha>\\
|\beta>
\end{array}
\right)\]
Where, $U^{-1}$ is given by
 \[
   U^{-1}=
  \left( {\begin{array}{cc}
   \cos\theta & \sin\theta \\       -\sin\theta & \cos\theta \      \end{array} } \right)
\]
So we have shown here the possibility that these states might be getting mixed up with each other to yield mixed states according to \cite{p1,p2,p3}

 \begin{equation}\label{eq:7}
|P_J\rangle=\sqrt{\frac{2}{3}}|\alpha \rangle + \sqrt{\frac{1}{3}}|\beta\rangle
\end{equation}
 \begin{equation}\label{eq:7}
|P'_J\rangle=-\sqrt{\frac{1}{3}}|\alpha  \rangle + \sqrt{\frac{2}{3}}|\beta\rangle
\end{equation}

Where, $|\alpha \rangle$ and $|\beta \rangle$ are same parity states.  The $|{P_J}'>$ and $|{P_J}> $ are the lower and higher eigen states respectively as given in Ref. \cite{Kenji}. For a finite mixing angle (or mixing probability ≤ 1) the masses of the $|{P_J}'>$ and $|{P_J}> $  states will lie only between the masses of the $|\alpha>$ and $|\beta> $ states.  Accordingly, we get mixed states at 10.914 GeV and 10.898 GeV for mixing of two states $^1P_1$ (10.931 GeV) and $^1P_1$ (10.882 GeV). Similarly for the mixing of $^1P_1$ (10.931) and $^5P_1$ (10.853), we obtained states at 10.905 GeV and 10.879 GeV and for that for the mixing of $^1P_1$ (10.882) and $^5P_1$ (10.853) states, we obtained mixed states at 10.871GeV and 10.862 GeV. In the same way we obtained mixed states for other combinations also. These mixed states are listed in Table 6. The masses of $1^{--}$ mixed states lie very close to the 10.890 resonance. We have also computed the leptonic and hadronic and radiative decay widths for these mixed states.
\begin{figure*}
\begin{center}
\resizebox{0.65\textwidth}{!}{%
  \includegraphics{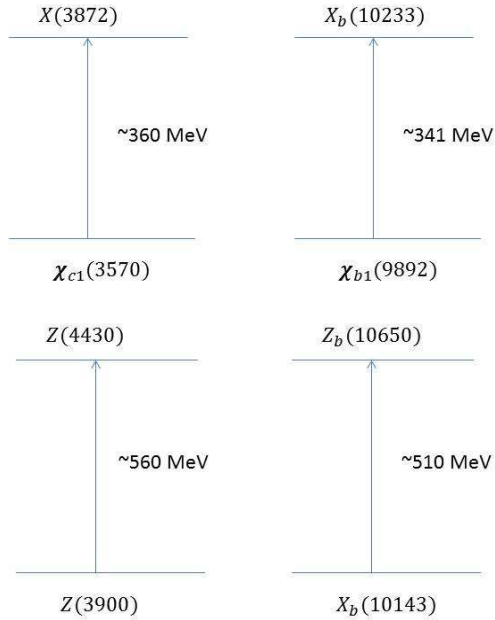}
}
%\vspace{5cm}       % Give the correct figure height in cm
\caption{ Bottom tetraqurk states analogue of charm tetraquark states in the mass region of interest.}
\label{fig:2}
       % Give a unique label
\end{center}
\end{figure*}
%%%%%%%%%%%%%%%%%%%%%%%%%%%%%%%%%%%%%%%%%%%%%%%%%%%%%%%%%%%%%%%%%%%%%%%%%%%%%%%%%%%%%%%%%%%%%%%%%%%%%%%%%%%%%%%
\section{$Y_b(10890)$ state and its decay properties}

The prominent exotic state $Y_b(10890)$ with $ J^{PC} = 1^{--}$ was first observed by the Belle collaboration
\cite{KF,I} and to date, it remains to be confirmed by independent experiments. The anomalously large
production cross sections for $ e^+e^- \rightarrow \Upsilon(1S;2S;3S)\pi^+\pi^-$ measured at $\Upsilon(5S)$ was not  in good agreement with the line shape and production rates for the conventional $b\bar b$  $\Upsilon(5S)$) state. An important issue is whether the puzzling events seen by Belle stem from the decays of the $\Upsilon(5S)$ or from another particle $Y_b$ having a
mass close enough to the mass of the $\Upsilon(5S)$. This results motivated theorists to resolve the puzzling features of this peak which lies approximately at mass 10.890 GeV. Currently, there are two competing theoretical explanations: the tetraquark interpretation on the one hand \cite{ali,A1,A2,A3} and the re-scattering model \cite{meng} on the other. The tetraquark model can explain the enhancement and the resonant structure via Zweig allowed decay processes and coupling to intermediate states while the re-scattering model is based on the decay $\Upsilon(5S))\rightarrow B^*{\bar B^*}$ and a subsequent recombination of the B mesons. A detailed study on this state is available in literature \cite{ali,A1,AA}.
The state $\Upsilon(10890)$ is usually referred as the $\Upsilon(5S)$, since its mass is close to the mass of the 5S state predicted by potential models. However, a different proposal has been put forwarded by the authors of Ref. \cite{ali}, in which they call this state as $Y_b$ and this state is being a P-wave tetraquark analogous to the $Y(4260)$, though the current experimental situation regarding the peak about 10.890 GeV is still debatable.\\
However, here we found three  vector states with $J^{PC} = 1^{--}$ whose mass is around 10.890 GeV i.e. 10.882GeV, 10.853 GeV and 10.931 GeV. We also found $1^{--}$ mixed $^1P_1$ states lie at 10.914 and 10.898 GeV. To resolve the $Y_b$ further we have calculated the di-electronic, hadronic and radiative decay widths of these states.

\subsection{Leptonic decay width of $J^{PC} = 1^{--}$ state}
\label{sec:3.1}

In the conventional $b\bar b$ systems, the decay widths are determined by the wave functions at the origin for ground state while for the P waves the derivation of these wave functions at the origin are used. We have used the same Van-Royen-Weisskopf formula but with a slight modification. Since tetraquark size is larger than that of quarkonia,to take into account larger size of tetraquark, we have modified wavefunctions by including a quantity $\sigma$, size parameter whose value varies from $\sigma\in[\frac{1}{2},\frac{\sqrt3}{2}]$ \cite{kvalue}. These tetraquarks wave functions will affect the decay amplitudes and thereby influencing the decay rates.
The partial electronic decay widths $\Gamma_{ee[bu]}$  and $\Gamma_{ee[bd]}$ of the tetraquark states $Y_{bu}$ and $Y_{bd}$ made up of diqaurks and antidiquarks (for up quark and down quark respectively) are given by the well known Van Royen Weiss-kopf formula for P waves\cite{ahmed}

\begin{equation}\label{eq:10}
\Gamma(Y_{[bu]/[bd]}\longrightarrow{e^+e^-})=\frac{24\alpha^2 <e_Q>^2}{M_{{{Y}^4_b}}}\sigma^2 |R'_{11}(0)|^2
%\Gamma(Y_{[bu]/[bd]}\longrightarrow}{e^+e^-})=\frac{24\alpha^2 {|Q_{[bu]/[bd]}|^2 |R'_{0}|^2}{M{_{Y_b]}^4}]}
\end{equation}

Here, $\alpha$ is the fine structure coupling constant and $\sigma<1$ and $<e_Q>$ is the effective charge of $Qq$ diquark  system given by\cite{Parmar}\\

 \begin{equation}\label{eq:11}
<e_Q>=|\frac{m_Q e_q-m_q e_Q}{m_Q+m_q}|
\end{equation}

 For computing the leptonic decay width, we have employed the numerically
obtained radial solutions while the authors\cite{ahmed,ali} have used value calculated by using $Q\bar Q$-onia package\cite{package} giving $|R'_{11}(0)|^2=2.067 GeV^5$. Our calculated results for leptonic decay widths for $Y_{bu}$ and $Y_{bd}$ are shown in Table 7 with available theoretical data. Since all the vector $ 1^{--}$ states are P-waves, the value of $R'(0)$  will not change as the masses of the diquarks remain the same. Hence, value of leptonic decay width does not change significantly as it only varies with the mass. However in the case of mixed states the contributions from the radial wavefnctions will be noticeable.

\subsection{Hadronic decay width of $J^{PC} = 1^{--}$ state}
\label{sec:3.2}
In this section, we have studied the hadronic decay of the $1^{--}$ P wave $Y_b(10890)$ state. We discuss the two body hadronic decays i.e. $Y_b(q)\rightarrow B_q^*(k)\bar B_q^*(l)$.These are zweig allowed processes and involve essentially the quark rearrangements. For calculating domiant two body hadronic decay widths of the $1^{--}$  $Y_b(10890)$ state, the vertices are given as \cite{peskin}

\begin{eqnarray}\label{eq:11}
Y_b\longrightarrow B\bar B &=&F(k^\mu-l^\nu) \nonumber\\
Y_b\longrightarrow B\bar B^* &=&\frac{F}{M}\epsilon^{\mu\nu\rho\sigma}k_\rho l_\sigma \nonumber\\
Y_b\longrightarrow B^*\bar B^* &=&F(g^{\mu\rho}(q+l)^\nu-g^{\mu\nu}(k+q)^\rho+g^{\rho\nu}(q+k)^\mu) \nonumber
\end{eqnarray}

and the corresponding decay widths are given by\\ \\
\small
\begin{equation}\label{eq:12}
\Gamma(Y_b\longrightarrow B\bar B) =\frac{F^2|\overrightarrow{k}|^3}{2M^2\pi}
\end{equation}
\normalsize
\small
\begin{equation}\label{eq:12}
\Gamma(Y_b\longrightarrow B\bar B^*)=\frac{F^2|\overrightarrow{k}|^3}{4M^2\pi}
\end{equation}
\normalsize
\small
\begin{equation}\label{eq:12}
\Gamma(Y_b\longrightarrow B^*\bar B^* ) =\frac{F^2|\overrightarrow{k}|^3(48|\overrightarrow{k}|^4-104M^2|\overrightarrow{k}|^2+27M^4)}{2\pi(M^3-4|\overrightarrow{k}|^2M)^2}
\end{equation}
\normalsize
Here $|\overrightarrow{k}|$ is the center of mass momentum given by

\begin{equation}\label{eq:11}
|\overrightarrow{k}|=\frac{\sqrt{M^2-(M_k+M_l)^2}\sqrt{M^2-(M_k+M_l)^2}}{2M}
%\sqrt{M^2-(M_k-M_l)^2}
\end{equation}

Where, M is the mass of the decaying particle and $M_k$, $M_l$ are the masses of the decay products.
The decay constant F is the non-perturbative quantity and to evaluate it is the beyond the scope in our approximation. We adopted the same approach used in \cite{ali,abdur}and estimate them using the known two body decays of $\Upsilon(5S)$ which are described by the same vertices as given in \cite{peskin}. To extract the value of F and $|\overrightarrow{k}|$, we have used the values of decay widths for the decays $\Upsilon(5S)\rightarrow B_q(k)\bar B_q(l),B_q(k)\bar B_q^*(l),B_q^*(k)\bar B_q^*(l)$ from Particle data group\cite{pdg2014}. The extracted value of the F and $|\overrightarrow{k}|$ are shown in the Table 8 along with the decay width results. To take into account different hadronic size of tetraquark we have included a quantity $\sigma$ that already discussed earlier. The results for hadronic decay widths are shown in table 5 which differ from the corresponding PDG\cite{pdg2014} values of $\Upsilon(5S)$. Out of these three P wave states, computed value of hadronic decay width for $Y_b(10853)$ is of the order of 50 MeV as against the PDG value of $110\pm13$ MeV and consistent with the BELLE measurements. For other two states we are getting more higher values than they actually should have. So out of these three states, we predict only the state with mass 10.853 GeV as a $Y_b(10890)$ state.

%%%%%%%%%%%%%%%%%%%%%%%%%%%%%%%%%%%%%%%%%%%%%%%%%%%%%%%%%%%%%%%%%%%%%%%%%%%%%%%%%%%%%%%%%%%%%%%%%%%%%%%%%%%%%%%%%%%%%%%%%%%%%%%%%

%%%%%%%%%%%%%%%%%%%%%%%%%%%%%%%%%%%%%%%%%%%%%%%%%%%%%%%%%%%%%%%%%%%%%%%%%%%%%%%%%%%%%%%%%%%%%%%%%%%%%%%%%%%%%%%%%%%%%%%%%%%%%%
\begin{table*}
\begin{center}
\caption{Mixed P wave states(in GeV))}. \label{tab1}
\begin{tabular}{cccc}
\hline\\
$J^{PC}$	&	State	&	Mixed State	\\
\hline

\hline
$1^{--}$	&	$^1P_1$(10.931)	&	10.914($P_J$)	\\
$1^{--}$	&	$^1P_1$(10.882)	&	10.898($P'_J$)	\\
	&		&		\\
$1^{--}$	&	$^1P_1$(10.931)	&	10.905($P_J$)	\\
$1^{--}$	&	$^5P_1$(10.853)	&	10.879($P'_J$)	\\
	&		&		\\
$1^{--}$	&	$^1P_1$(10.882)	&	10.862($P_J$)	\\
$1^{--}$	&	$^5P_1$(10.853)	&	10.871($P'_J$)	\\
&		&		\\
$0^{-+}$	&	$^3P_0$(10.883)	&	10.876($P_J$)	\\
$0^{-+}$	&	$^3P_0$(10.862)	&	10.868($P'_J$)	\\

\hline

\hline

\end{tabular}

\end{center}

\end{table*}
\begin{table*}

\begin{center}
\caption{Di-leptonic decay widths (in keV))}. \label{tab1}

\begin{tabular}{ccccc}
\hline

State	&	$\Gamma_{ee[bl]}$	&	$\Gamma_{ee[bh]}$	\\
\hline
$Y_b(10882)$	&	0.0251	&	0.123	\\
	&		&		\\
$Y_b(10853)$	&	0.0254	&	0.125	\\
	&		&		\\
$Y_b(10931)$	&	0.0246	&	0.121	\\
&		&		\\
$Y(10914)$(Mixed state)	&	0.02485	&	0.122	\\
	&		&		\\

$Y(10898)$(Mixed state)	&	0.02499	&	0.1229	\\
	&		&		\\
$Y(10905)$(Mixed state)	&	0.0249	&	0.1226	\\
	&		&		\\
$Y(10879)$(Mixed state)	&	0.02517	&	0.1238	\\
	&		&		\\
$Y(10862)$(Mixed state)	&	0.02532	&	0.1245	\\
	&		&		\\
$Y(10871)$(Mixed state)	&	0.02524	&	0.1241	\\
	&		&		\\
Others	&	$0.09\pm0.03$\cite{ali}	&	$0.08\pm0.03$\cite{ali}	\\
	&	&$0.12$\cite{abdur}	&		\\

\hline

\end{tabular}
\end{center}
\end{table*}	
%%%%%%%%%%%%%%%%%%%%%%%%%%%%%%%%%%%%%%%%%%%%%%%%%%%%%%%%%%%%%%%%%%%%%%%
\subsection{Radiative decay width of $J^{PC} = 1^{--}$ state}
\label{sec:3.13}
We study the radiative decays of these states using the idea of Vector Meson Dominance (VMD) which describe interactions
between photons and hadronic matter \cite{vmd} and we hope that this will increase an insight about these tetraquark states.
The transition matrix element for radiative decay of $Y_b\rightarrow\chi_b+\gamma$ is given with the use of VMD \\
\begin{equation}\label{eq:12}
<\chi_b\mid\gamma>=<\gamma\mid\rho>\frac{1}{{m_\rho}^2}<\chi_b \rho \mid Y_b>
%\alpha_s(\mu^2)=\frac{4\pi}{(11-\frac{2}{3}n_f)(\ln\frac{\mu^2}{\Lambda^2+{M^2_B}})}
\end{equation}
and the decay width is given by
\begin{equation}\label{eq:13}
\Gamma(Y_b\rightarrow\chi_b+\gamma)=2|A^2|(\frac{f_\rho}{{m_\rho}^2})^2\frac{1}{8\pi {M_{Y_b}}^2}\frac{(\lambda)^\frac{1}{2}}{2M_{Y_b}}
%\alpha_s(\mu^2)=\frac{4\pi}{(11-\frac{2}{3}n_f)(\ln\frac{\mu^2}{\Lambda^2+{M^2_B}})}
\end{equation}
Where, $\lambda$ is the center of mass momentum and $f_\rho=0.152 GeV^2$\cite{value}. Similarly, we have computed radiative decay $Y_b\rightarrow\eta_b+\gamma$. The present results are shown in the Table 9 with available theoretical data. There is no experimental data available for the radiative decay of $Y_b(10890)$ and we look forward to see the experimental support in favour of our predictions.

\begin{table*}
%\small
\begin{center}
\caption{ Reduced partial hadronic decay widths and reduced total decay widths(in keV), the extracted value of the coupling constant F and the centre of mass momentum $|{\vec k}|$ .} \label{tab1}
%\scalebox{0.98}{
\begin{tabular}{ccccccc}
\hline

State	&	Decay mode	&	F	&	$  |\overrightarrow{k} |$	&	$\Gamma$&	 $\frac{\Gamma}{ \sigma^2}$	& $\frac{\Gamma_{tot}} {\sigma^2}$\\
\hline\\
	&	$Y_{bq}\rightarrow B\bar B$	&	1.35	&	1.31	&	5.500&	6.790&71.89	\\
$Y_b(10882)$	&	$Y_{bq}\rightarrow B{\bar B}^*$	&	3.12	&	1.22	&	11.87&	14.66&	\\
	&	$Y_{bq}\rightarrow{B^*\bar B}^*$	&	0.92	&	1.11	&	40.86&	50.44	&\\
	&		&		&		&			\\
	&	$Y_{bq}\rightarrow B\bar B$	&	1.35	&	1.25	&	4.800&	5.930&	56.08\\
$Y_b(10853)$	&	$Y_{bq}\rightarrow B{\bar B}^*$	&	3.12	&	1.15	&	10.00	&12.34&	\\
	&	$Y_{bq}\rightarrow{B^*\bar B}^*$	&	0.92	&	1.04	&	30.63	&37.81&	\\
	&		&		&		&			\\
	&	$Y_{bq}\rightarrow B\bar B$	&	1.35	&	1.41	&	6.800	&8.400&	97.45\\
$Y_b(10931)$	&	$Y_{bq}\rightarrow B{\bar B}^*$	&	3.12	&	1.32	&	14.91&	18.40&	\\
		&$Y_{bq}\rightarrow B^*{\bar B}^*$		&0.92	&	1.23		&57.23&	70.65&	\\ \\

Mixed P states&		&		&		&			\\
	&	$Y_{bq}\rightarrow B\bar B$	&	1.35	&	1.381	&	6.399&	7.900&89.09	\\
$Y(10914)$	&	$Y_{bq}\rightarrow B{\bar B}^*$	&	3.12	&	1.29	&	13.96&	17.23&	\\
	&	$Y_{bq}\rightarrow{B^*\bar B}^*$	&	0.92	&	1.19	&	51.81&	63.96	&\\
	&		&		&		&			\\
	&	$Y_{bq}\rightarrow B\bar B$	&	1.35	&	1.33	&	5.745&	7.098&	76.72\\
$Y(10898)$	&	$Y_{bq}\rightarrow B{\bar B}^*$	&	3.12	&	1.23	&	12.13	&14.98&	\\
	&	$Y_{bq}\rightarrow{B^*\bar B}^*$	&	0.92	&	1.13	&	44.27	&54.65&	\\
	&		&		&		&			\\
	&	$Y_{bq}\rightarrow B\bar B$	&	1.35	&	1.41	&	6.135	&7.574&	84.83\\
$Y(10905)$	&	$Y_{bq}\rightarrow B{\bar B}^*$	&	3.12	&	1.32	&	13.34&	16.47&	\\
		&$Y_{bq}\rightarrow B^*{\bar B}^*$		&0.92	&	1.23		&49.24&	60.79&	\\
&		&		&		&			\\
	&	$Y_{bq}\rightarrow B\bar B$	&	1.35	&	1.41	&	5.509	&6.802&	73.01\\
$Y(10879)$	&	$Y_{bq}\rightarrow B{\bar B}^*$	&	3.12	&	1.32	&	11.59&	14.31&	\\
		&$Y_{bq}\rightarrow B^*{\bar B}^*$		&0.92	&	1.23		&42.06&	51.90&	\\ \\
&	$Y_{bq}\rightarrow B\bar B$	&	1.35	&	1.27	&	5.035	&6.217&	64.35\\
$Y(10862)$	&	$Y_{bq}\rightarrow B{\bar B}^*$	&	3.12	&	1.17	&	10.51&	12.98&	\\
		&$Y_{bq}\rightarrow B^*{\bar B}^*$		&0.92	&	1.06		&36.58&	45.17&	\\ \\
&	$Y_{bq}\rightarrow B\bar B$	&	1.35	&	1.29	&	5.268	&6.504&	69.26\\
$Y(10871)$	&	$Y_{bq}\rightarrow B{\bar B}^*$	&	3.12	&	1.19	&	11.04&	13.63&	\\
		&$Y_{bq}\rightarrow B^*{\bar B}^*$		&0.92	&	1.09		&39.80&	49.14&	\\

\hline

\end{tabular}
%}
\end{center}

\end{table*}
%%%%%%%%%%%%%%%%%%%%%%%%%%%%%%%%%%%%%%%%%%%%%%%%%%%%%%%%%%%%%%%%%%%%%%%%%%%%%%%%%%%%%%%%%%%%%%%%%%%%%%%%%%%%%%%%%%%%%%%%%
\begin{table*}
\begin{center}
\caption{Radiative decay widths (in keV))}. \label{tab1}
\begin{tabular}{ccc|cccc}
\hline
State&$\Gamma\rightarrow\chi_b+\gamma$ &	$\frac{\Gamma \rightarrow \chi_b+\gamma}{\Gamma \rightarrow\Upsilon+\pi^+ +\pi^-}$&$\Gamma\rightarrow\eta_b+\gamma$ &	$\frac{\Gamma \rightarrow \eta_b+\gamma}{\Gamma \rightarrow\Upsilon+\pi^+ +\pi^-}$								\\
	\hline
$Y_b(10882)$	&	0.173	&	0.293	& 0.247	&	0.418							\\
	&		&									\\
$Y_b(10853)$	&	0.169	&	0.286	& 0.243	&	0.413							\\
	&		&									\\
$Y_b(10931)$	&	0.179	&	0.304	& 0.252	&	0.427							\\
&		&		&		&			\\
Mixed P states&		&		&		&			\\
$Y(10914)$	&	0.177	&	0.300	& 0.250 &0.424							\\
&		&		&		&			\\	
$Y(10898)$	&	0.175	&	0.296	& 0.248 &0.421							\\
&		&		&		&			\\	
$Y(10905)$	&	 0.176	&	0.298	& 0.249 &0.422							\\
&		&		&		&			\\	
$Y(10879)$	&	0.172	&	0.292	& 0.246 &0.418							\\
	&		&		&		&			\\
$Y(10862)$	&	0.170	&	0.288	& 0.244 &0.415							\\
	&		&		&		&			\\									
$Y(10871)$	&	0.171	&	0.291	& 0.245 &0.416							\\
&		&		&		&			\\
Others\cite{abdur}          &      -     &    0.3 & - &    0.5              \\
\hline

\end{tabular}

\end{center}

\end{table*}		
%%%%%%%%%%%%%%%%%%%%%%%%%%%%%%%%%%%%%%%%%%%%%%%%%%%%%%%%%%%%%%%%%%%%%%%%%%%%%%%%%

%%%%%%%%%%%%%%%%%%%%%%%%%%%%%%%%%%%%%%%%%%%%%%%%%%%%%%%%%%%%%%%%%%%%%%%%%%%%%%%%%%%%%%%%%%%%%%%%%%%%%%%%%%%%%%%%%%%%%%%
\begin{table*}
\begin{center}
\caption{Interpretation of some $1^{st}$ radially excited states.} \label{tab8}
\begin{tabular}{cccc}
\hline\hline\\

$J^{PC}$	&	State		&		$1^{st}$ radial excitation	&Exp	 \\
\hline\\
$0^{++}$	&  $X_b$(10.143)	&	$10.650[Z_b(10650)]$&$10.652\pm0.0025$\cite{bonder}	 \\
$1^{+-}$	&  $X_b$(10.233)	&	$10.683[Z_b(10650)]$&	 \\
$1^{--}$	&  $Y_b$(10.853)	&	$11.095[Y_b(?)]$&	 \\
$1^{--}$	&  $Y_b$(10.882)	&	$11.108[Y_b(?)]$&	 \\
$1^{--}$	&  $Y_b$(10.931)	&	$11.151[Y_b(?)]$&	 \\
\hline\hline

\end{tabular}

\end{center}
\end{table*}
\section{Results and Discussions}
We have computed the mass spectra of hidden bottom four quark states in diquark-antidiquark
picture which are listed in table 1, 2, 3, 4 and 5. We have taken various combinations of the orbital and spin excitations to compute the mass spectra. The computed mass spectra are compared with other available theoretical results in figure 1. Apart from this we mainly have paid attention to $Y_b(10890)$ state and have computed leptonic, hadronic and radiative decay width of $Y_b$ which are listed in tables 7, 8 and 9 respectively. Apart from this, we have also done mixing of $1^{--}$ P waves which are also listed in the respective Tables. The core of the present study is that the color diquark is handled as a constituent building block. We predicted some of the bottom tetraquark states as counterpart in the charm sector. It is necessary to highlight that the observation of the bottom counterparts to the new anomalous charmonium-like states is very important since it will allow to distinguish between different theoretical descriptions of these states. In this viewpoint, it would be also valuable to look for the analogue in the bottom sector as states related by heavy quark symmetry may have universal behaviours. The predicted bottom counterparts are shown in the Fig. 2 for better understanding.
In the present study, we have noticed that mass difference between predicted  $X_b(10233)$ and ${\chi_{b_1}}(9892)$

\begin{equation}\label{eq:1}
M_{X_b}-M_{\chi_{b_1}}  \sim 341 MeV
\end{equation}

which of the same order of magnitude of the mass difference  between $X(3872)$ and $\chi_{c_1}(3510)$ of the charm sector
\begin{equation}\label{eq:2}
M_X-M_{\chi_{c_1}}  \sim 360 MeV
\end{equation}
This kind of similarity between charm and bottom sector is very interesting.
We found that mass difference between $X_b(10143)$ and its first radially excited state $X_b(10650)$  states is $ \sim 510 MeV$
as similar to charmonia which is about 590 MeV. In the same way, we found that mass difference between $X_b(10233)$ and its first radially excited state $X_b(10683)$  states is $ \sim 450 MeV$.
So by taking the the evidence from these results, we can say that 4-quark state in the bottom sector analogous to charm sector should exist. We have predicted some of the radially excited states which are listed in Table 10. Accordingly, we predicted $Z_b(10650)$ state as the first radial excitation of either $X_b(10143)$ ($0^{++}$) state or $X_b(10233)$($1^{+-}$) state. The authors of  Ref. \cite{Guo} studied the masses of the S-wave  $[bq][\bar  b\bar q]$  tetraquark states  with the inclusion of chromomagnetic interaction and they predicted  the lowest $[bq][\bar  b\bar q]$ tetraquark state appears at 10.167 GeV. This result is consistent with the results of  Ref. \cite{cui} where, using the color-magnetic interaction with the flavor symmetry breaking corrections, the $[bq][\bar  b\bar q]$  tetraquark states were predicted to be around $10.2-10.3 $ GeV. The same results are found by authors of  Ref.  [36, 37] where they have used the QCD sum rule approach for the computation of mass pectra of $[bq][\bar  b\bar q]$  tetraquark state. The authors of Ref. \cite{1} have used different tetraquark $[bq][\bar  b\bar q]$ currents and they have obtained $M_{X_b} = (10220\pm100)$ MeV, which is in complete agreement with the result of Ref. \cite{2}. These predictions of $X_b$ state and its production rates in hadron-hadron collisions have indicated a promising prospect to find the $X_b$ at hadron collider in particular the LHC and we suggest our experimental colleagues to perform an analysis. Such attempt will likely lead to the discovery of the Xb and thus enrich the list of exotic hadron states in the heavy bottom sector. An observation of the $X_b$ will provide a deeper insight into the exotic hadron spectroscopy and is helpful to unravel the nature of the states connected by the heavy quark symmetry. Similarly, there exist other radial excited states in the region 11.095-11.151 GeV corresponding to 2P states. We look forward to see experimental search for these states. The authors in Refs \cite{POS,Dong} have predicted $Z_b(10650)$ state as di-mesonic molecular state in the ground state. From our present study, we suggest that if $Z_b$ states are diquark -diantiquark states then they are not the ground state of  bottomonium-like four quark state but the first radially excited state of its ground state which lies in the $10.100-10.300 GeV$ which is in agreement with the results reported by the authors of Ref \cite{FSN}. The same presumption was made by authors of Ref \cite{LFAD,sam} to explain $Z(4430)$ state as an excitation of state $Z_c(3900)/Z_c(3885)$ in the charm sector. So in conjecture with this, our prediction regarding $Z_b(10650)$ state is just a straightforward extension to beauty sector and we observe that the $Z_b(10650)$ is also a radially excited state of still unmeasured $X_b$ state just like that of authors of Ref \cite{FSN} who predicted $Z_b(10610)$ state as the the radial excitation of $X_b(10100)$ such that the mass difference is $M_{Z_b}(10650)-M_{X_b(10143)}\sim 510 MeV$ which is very close to mass difference between $\Upsilon(2S)-\Upsilon(1S) = 560 MeV$.
 To have a clear-cut picture about the discussion made regarding the bottom exotic states, the above discussed exotic states are displayed in the Fig. 2 with analogous states at the charm sector.

The comparison between the bottom tetraquark states and charm tetraquark states accentuates the resemblance between presumptions made in the present study, namely the existence of a $X_b(10143)$ as a ground state of $Z_b(10650)$
and presumption related to existence of ground state of $Z(4430)$ made in Refs\cite{LFAD,sam}. The presumption of bottom tetraquark states analogous to charm spectra should stimulate searches for these states in both the beauty and  the charm sector within the mass range around $10100-10300 MeV$ and $3500-3870 MeV$ respectively. The searching of these states would be not only able to find unobserved state shown in Fig. 1 but also be able to detect many more prominent states in these mass range. As $Y_b(10890)$ state with quantum number $1^{--}$ is of our keen of interest, in this study we have predicted three P wave $1^{--}$ states in the mass region around 10.850-10.931 GeV. We have observe that P wave state with mass 10.853 GeV as the  $Y_b$ state. The calculated partial electronic decay widths for P wave $Y_b$ is about 0.03-0.12 keV which is in agreement with the available experiment data \cite{babar} and other theoretical predictions \cite{ali,abdur}. Our present calculation show that the leptonic width of $Y_b$ is much lower than that of the width of conventional state $\Upsilon(5S)(0.31\pm0.07keV)$\cite{pdg2014}. From this we can say that  $\Upsilon(10890)$ peak is different from the $\Upsilon(5S)$ and possibly may be the $Y_b(10890)$ only. We have also computed the two body hadronic decays of $Y_b$. The total hadronic decay width is of the order of 50 MeV which is lower than the total decay width of $\Upsilon(5S)=110MeV$ state. So this narrow width state $Y_b(10890)$ can be tetraqurk state only rather than being the conventional $b\bar b$ state. We have also computed the radiative decay widths of $Y_b$, but due to lack of experimental results we can not make any concrete conclusion here. These results can be guidelines for future studies. In the absence of experimental data, we can't make any conclusion regarding mixing of P wave states but we expect our results could be helpful to understand the structure of these states. Our computed masses of $1^{--}$ mixed states i.e. $^1P_1$ and $^5P_1$ states lie very close to $Y(10890)$ state by at most an order of $\pm20$ MeV. So we look forward to see the experimental search for these states with very high precision as these states are very closely spaced. The experiments should have in principle the sensitivity to detect and also to explore the nature of such near-lying states. The present study of mixing is an attempt to signify its importance to further resolve mystery of $Y_b(10890)$. If  the status of $Y_b(10890)$ is confirmed then it will be a major step in the direction of testing the models and provide theorists with vital input to present a credible explanation of this new form of hadrons.



\end{document}